\documentclass[9pt,twocolumn,oneside]{extarticle}
\usepackage{authblk}
\usepackage{graphicx}
\usepackage{cite}
\usepackage{hyperref}
\title{\vskip-5em\textbf{Enhancement of terahertz emission during single-color filamentation by chirping laser pulse}}

\author[1,*]{Georgy Rizaev}
\author[1]{Dmitrii Pushkarev}
\author[1,2]{Maximilian Levus}
\author[2]{Nadezhda Vrublevskaya}
\author[1]{Leonid Seleznev}

\affil[1]{Lebedev Physical Institute of RAS, 53 Leninskiy pr., Moscow 119991, Russia}
\affil[2]{Faculty of Physics, Lomonosov Moscow State University, Leninskie Gory, Moscow 119991, Russia}

\affil[*]{rizaev@lebedev.ru}
\date{}

\begin{document}

\begin{abstract}
An experimental study of laser pulse duration influence on the terahertz emission during single-color filamentation is carried out. It is shown that for each terahertz frequency there is an optimal laser pulse duration providing maximal generation at constant pulse energy. It is demonstrated that longer pulses are required for stronger low-frequency terahertz emission, thus despite considerable laser peak power decreasing the terahertz radiation yield can be increased by more than 3~times.
\end{abstract}

\maketitle

Terahertz radiation (0.1 –- 10~THz) has low quantum energy, and therefore has little effect on the propagation medium. This feature of terahertz radiation opens up broad prospects for its use in such areas as medicine~\cite{amini2021}, terahertz spectroscopy~\cite{baxter2011}, security systems~\cite{federici2005}, industrial control~\cite{wietzke2009} and other~\cite{leitenstorfer2023}. There are many ways to generate terahertz radiation, for example, using antennas, converting laser pulses in nonlinear crystals, etc.~\cite{lewis2014, fulop2020}. One of the simplest ways to generate broadband terahertz radiation is femtosecond laser filamentation~\cite{sun2022}. Energy, propagation angles and spectral composition of the generated terahertz radiation are important parameters for possible applications. Some studies report how the energy~\cite{jahangiri2012, mokrousova2020}, focusing conditions~\cite{mokrousova2020, rizaev2023}, and polarization~\cite{rizaev2022} of the laser pulse affect the parameters of terahertz emission. As for laser pulse duration, there are several recent papers on two-color filamentation with numerical~\cite{nguyen2018, saeed2023} or both experimental and theoretical~\cite{zhang2018, flender2019, mou2023, yu2022, xu2023, martinez2024} study of the laser pulse duration effect on terahertz emission. As the main conclusion from these studies, the pulse chirp allows control overlap and coincidence of the relative phase of fundamental and second harmonics pulses, thereby increasing the terahertz emission output. However, this effect is obviously not applicable to the single-color filamentation. We managed to find only one study of the effect of laser pulse duration on terahertz emission during single-color filamentation~\cite{akturk2007}. In this experiments on filamentation in xenon, the initial laser pulse was stretched by almost three times from $\sim$40 to $\sim$120~fs by clipping the bandwidth at the compressor part of CPA system. The laser pulse power was kept constant (the energy increased with duration stretching) to maintain conditions for filamentation. Terahertz radiation was measured by a narrow-frequency heterodyne detector at a frequency of 100~GHz. The authors obtained a quadratic growth in the  terahertz radiation energy with pulse duration increase, that can be attributed to the pulse energy rise. The effect of laser pulse energy on the terahertz radiation yield was studied earlier in~\cite{amico2007}, where a quadratic dependence was also observed. Thus, the authors implicitly state that the energy of terahertz radiation does not depend on the laser pulse duration. In a theoretical study~\cite{sprangle2004} it was shown that the duration of the terahertz pulse in air should be a few hundred femtoseconds, i.e., the laser pulse duration plus the plasma electron collision time. The electron collision time was estimated as 200~fs. From this point of view, the terahertz signal recorded in~\cite{akturk2007} has a duration of about 10~ps (at 100~GHz), which is much longer than 240 or 320~fs (the original duration is from 40 to 120~fs + 200~fs electron collision time). In other words, the terahertz signal in the experiments was measured at the frequency far from the maximum of the emitted terahertz spectrum, so these measurements did not allow to reveal the dependence on the laser pulse duration. Therefore, the goal of our work is to study experimentally the effect of laser pulse duration, varying over a wide range, on the energy of terahertz radiation generated at different frequencies during single-color filamentation in air.

\begin{figure}[h]
\centering
\includegraphics[width=\linewidth]{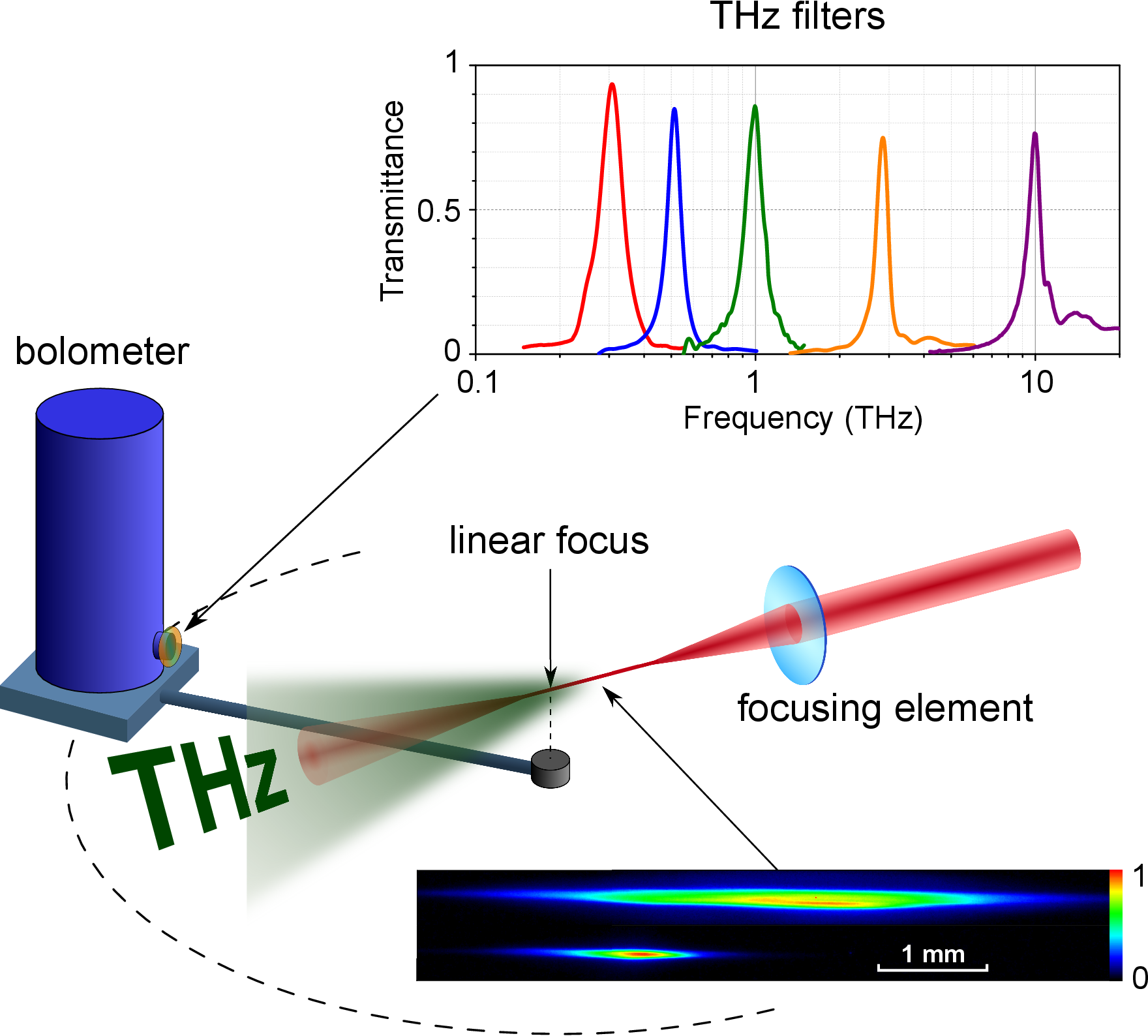}
\caption{Scheme of the experiment. The inset at the top shows the narrowband filters transmittance. The inset at the bottom shows plasma channel images at pulse durations of 90~fs (above) and 900~fs (below). Laser pulse propagates from right to left.}
\label{Fig1}
\end{figure}

In experiments we use a Ti:Sapphire laser system with pulse central wavelength of 750~nm and repetition rate of 10~Hz. The laser beam polarization is vertical. The minimal pulse duration is 90~fs (FWHM) and by moving the laser system compressor gratings we can stretch the pulse up to 900~fs, adding positive or negative chirp. After the laser system, the beam is focused and a plasma channel is formed near the geometric focus (Fig.~\ref{Fig1}). As a focusing element, we use a dielectric spherical mirror with focal length of 25~cm. Terahertz radiation generated in the filament plasma is detected with a superconducting hot-electron bolometer. To observe discrete terahertz frequency we place a narrow-band filter in front of the bolometer input window. The spectral transmittance of our filters is given in Fig.~\ref{Fig1}. To obtain the angular distribution of terahertz radiation in the horizontal plane, the bolometer is rotated around the geometric focus (Fig.~\ref{Fig1}).

At fixed laser pulse energy, the pulse stretching leads to the peak power decrease and, consequently, to a change in self-focusing conditions. To characterize the plasma channel parameters at different pulse durations, we observe the side image of the plasma channel, using a CCD with an objective lens. For example, plasma images obtained at pulse durations of 90~fs (top) and 900~fs (bottom) are shown in Fig.~\ref{Fig1}.

\begin{figure}[h!]
\centering
\includegraphics[width=0.95\linewidth]{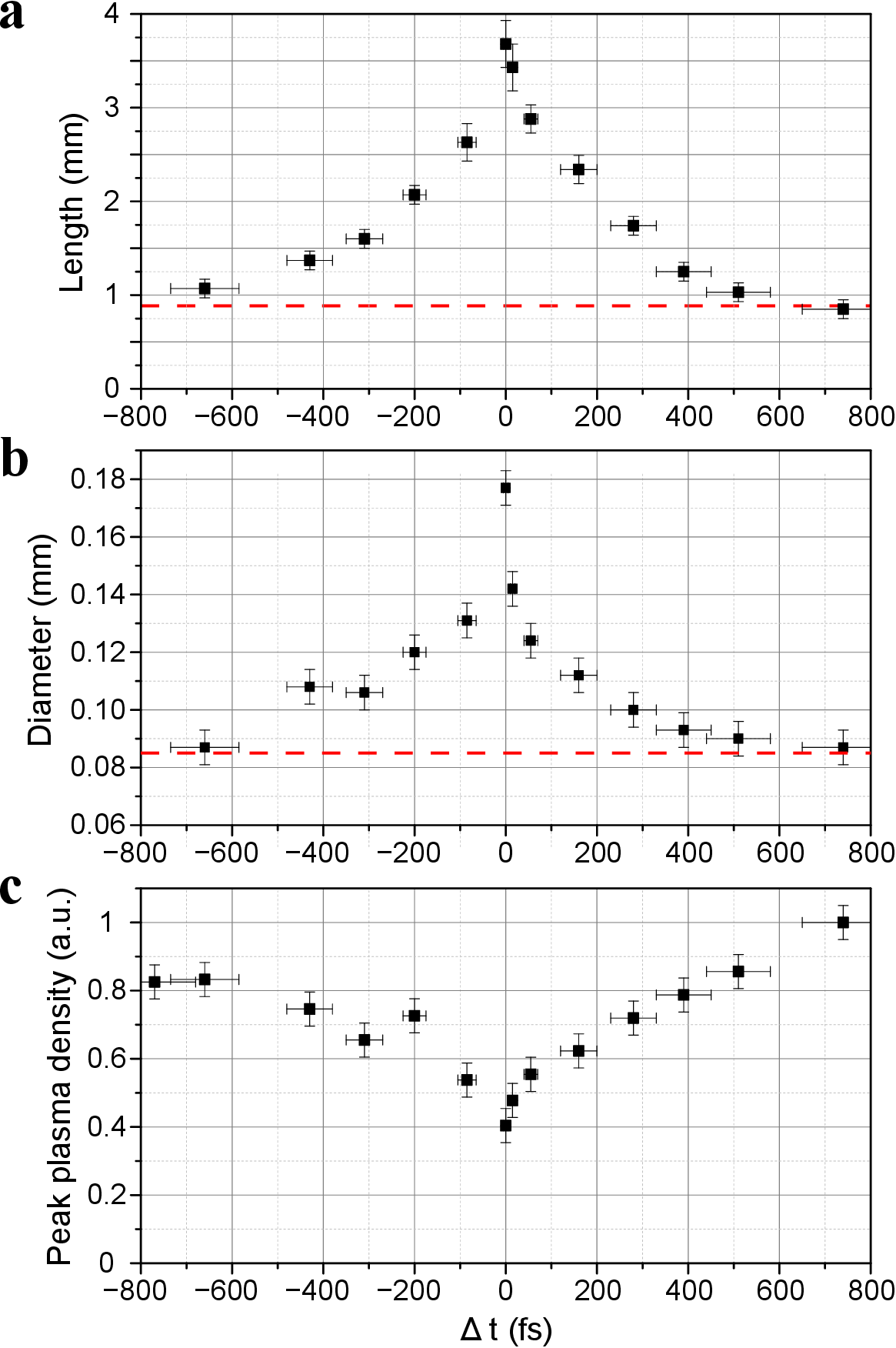}
\caption{Dependencies of plasma channel length (a), diameter (b) and peak plasma density (c) on the stretching of laser pulse. The red dashed line shows the corresponding dimensions for the Gaussian beam waist.}
\label{Fig2}
\end{figure}

Fig.~\ref{Fig2}a,b shows the dependencies of the plasma channel length and maximal diameter at the 1/e level on pulse stretch duration $\Delta$t. The value $\Delta$t = 0 corresponds to minimal pulse duration of 90~fs. To distinguish between positively and negatively chirped pulses, we indicate the corresponding sign for $\Delta$t, i.e. $\Delta$t < 0 does not mean pulse compression, but negatively chirped stretching. At $\Delta$t = 0 the peak power is the highest and due to significant contribution of self-focusing the plasma channel starts considerably before focal waist. In this case, the plasma channel length is maximal and reaches almost 4~mm (Fig.~\ref{Fig2}a). The pulse stretching leads to weaker self-focusing contribution into the plasma channel formation, and at durations of $\sim$600~fs ($\Delta$t$\sim$500~fs) the plasma is produced only on waist length scales (shown in the figure with the horizontal red dashed line). The plasma channel diameter has a similar behavior (Fig.~\ref{Fig2}b), i.e. greater pulse powers results in wider plasma channels, as in paper~\cite{theberge2006}.

For the visible spectral range recorded by the CCD the critical plasma density is about 10$^{21}$ cm$^{-3}$. The filament plasma channel density is much lower (about 10$^{17}$~cm$^{-3}$~\cite{couairon2007}) and the channel is transparent for plasma luminescence. Therefore the detected side luminescence signal is integrated over the channel width. Thus, in our experiments the peak plasma density is estimated as the maximal plasma luminescence signal divided by the plasma channel diameter. The laser pulse stretching from 90 to 900~fs results in almost twofold growth of the peak plasma density (Fig.~\ref{Fig2}c).

\begin{figure}[h!]
\centering
\includegraphics[width=0.85\linewidth]{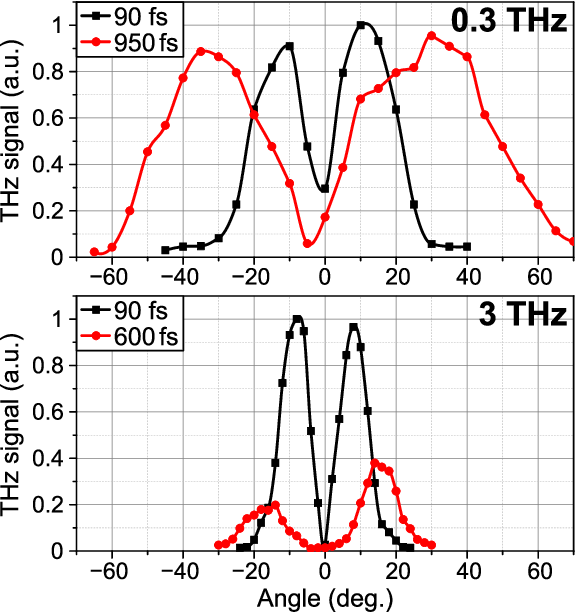}
\caption{Angular distributions of terahertz emission at frequencies of 0.3 and 3~THz, obtained at different laser pulse durations.}
\label{Fig3}
\end{figure}

The angular distribution of terahertz emission at different frequencies is measured for different laser pulse durations. For example, Fig.~\ref{Fig3} shows the distributions at frequencies of 0.3 and 3~THz at various pulse durations. In both cases, pulse stretching leads to increase in the propagation angles of terahertz emission. When the laser pulse is stretched, the terahertz signal peak amplitude does not change at low frequencies, but at high frequencies it drops significantly. It should be pointed out that with the long pulse the terahertz emission at 0.3~THz is measured at angles up to 70${^\circ}$ from the optical axis, so it would be problematic to collect the whole terahertz beam on the detector by optical methods.

\begin{figure}[h!]
\centering
\includegraphics[width=0.85\linewidth]{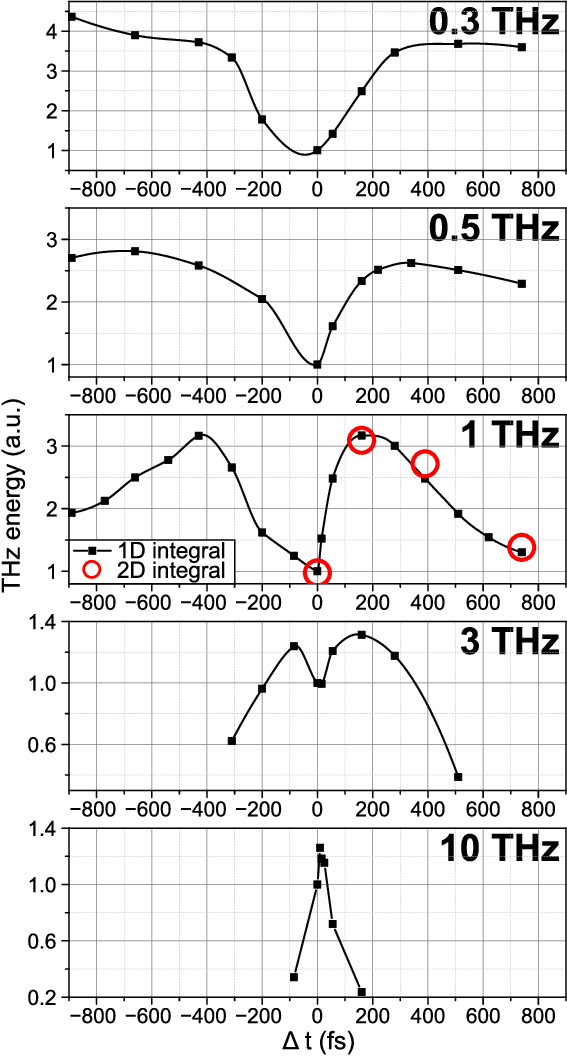}
\caption{Dependence of terahertz pulse energy on the stretching of laser pulse at different frequencies. The red circles indicate terahertz pulse energy obtained from two-dimensional pictures.}
\label{Fig4}
\end{figure}

To obtain the dependence of the terahertz pulse energy at different frequencies on laser pulse duration, the terahertz signals are integrated over one-dimensional angular distributions. However, the two-dimensional pattern of terahertz emission may be much different at various frequencies~\cite{nikolaeva2023}. Therefore, during the integration we take into account these spatial structures at different terahertz frequencies. The obtained dependencies are presented in Fig.~\ref{Fig4}.

To verify our integration method, we measure two-dimensional terahertz angular distributions for different pulse durations at frequency of 1~THz. At this frequency, the strongest deviation from axial symmetry was observed under our experimental conditions~\cite{rizaev2022, nikolaeva2023}. To obtain two-dimensional terahertz pattern we measure the distribution in the horizontal plane for different vertical angles. To change vertical angles we rotate the optical axis in the vertical plane by moving and adjusting the focusing mirror. At the same time, the position of the geometric focus is maintained at the same point. The  method for measuring two-dimensional terahertz distributions is described in detail in~\cite{rizaev2023}. Integrating signals over a two-dimensional distribution provides a more reliable estimation of the terahertz radiation energy. The obtained values are normalized and shown in Fig.~\ref{Fig4} with red circles. The circles size corresponds to the experimental error. The data obtained from two- and one-dimensional distributions coincide within the error, that allows to investigate the terahertz radiation energy basing on one-dimensional distributions.

Fig.~\ref{Fig4} shows the dependence of the terahertz pulse energy at different frequencies on the laser pulse stretch $\Delta$t. At all observed frequencies, except 10~THz, pulse stretching causes an increase in the terahertz emission energy. Then the terahertz yield reaches its maximum at a certain pulse duration for each frequency and begins to decrease with further pulse elongation. $\Delta$t corresponding to the maximum of terahertz generation declines with frequency increasing. At a frequency of 10~THz, a decrease is observed for any pulse stretching, hence the maximal terahertz generation corresponds to a shorter duration than of our unstretched pulse. To recap, for each terahertz frequency there is an optimal pulse duration providing the highest yield. Indeed, the timescale of ionisation and plasma interaction with the laser pulse depend on its duration. This timescale along with the characteristic plasma relaxation time determines the terahertz frequencies that would be predominately emitted by the plasma. As an interesting result, when the laser pulse peak power is reduced many times, the energy of low frequency ($\leq$1~THz) terahertz radiation increases by 3-4 times.

It seems the most obvious to associate the frequency of terahertz emission from a single-color filament with the plasma frequency. Under our conditions, when the laser pulse is stretched, the peak plasma density increases slightly (Fig.~\ref{Fig2}c), i.e. the plasma frequency increases also. So, the higher frequency terahertz components should be generated more efficiently. However, in our experiments, the pulse stretching leads to the increase in the low-frequency terahertz emission. Therefore, the plasma frequency is not directly related to the terahertz emission spectrum.

On the other hand, the characteristic terahertz frequency can be determined by the laser pulse duration and the electron collision frequency~\cite{sprangle2004}. Under our conditions, the electron-heavy particle (neutrals and ions) rate is about 1.5$\cdot$10$^{-13}$ s$^{-1}$~\cite{mokrousova2020}. The results of our experiments show the good agreement with the estimations for 3~THz. If we consider the optimal duration obtained in the experiment for a frequency of 1~THz and carry out similar estimations, the resulting frequency is 1.8—3~THz, i.e. this approach is not applicable over the entire terahertz frequency range.

The paper~\cite{amico2008} proposes another logical correlation between the laser pulse duration and the characteristic terahertz frequency. The laser pulse ionizes the air and pushes electrons aside by ponderomotive force and light pressure. After the pulse passes, the electrons return under the influence of the Coulomb force. If we match the electron current and the terahertz emission, then the laser pulse duration should correspond to the one-fourth of the terahertz period. From this approach, the optimal duration for 0.5~THz is about 500~fs. Comparing the maxima obtained in our experiments for positive and negative chirps, we obtain reasonable agreement with this estimation. In contrast, for 3~THz the model under consideration predicts optimal pulse duration of 80~fs, which is 2-3 times shorter than observed in the experiment. Thus, in order to find the correlation between the laser pulse duration and the characteristic terahertz frequency in wide spectral range, it is not enough to use simple models; it is necessary to carry out more complicated numerical simulation. The fact observed in our experiments that the maximum terahertz emission yield corresponds to different laser pulse duration for positive and negative chirps also cannot be explained within existing models.

In conclusion, we have experimentally studied the effect of laser pulse duration on the terahertz emission from the plasma of a single-color filament. The energy of the terahertz pulse at separate frequencies has been obtained by integrating the angular distribution, that allows to collect terahertz emission in a wide angular range. As a result, the maximal laser pulse power (minimal duration in our experiment of 90~fs) has been shown not to be optimal for terahertz generation at frequencies lower than 3~THz. With laser pulse stretching at a constant energy, the terahertz emission energy at low frequencies can be increased by several times.

%\begin{backmatter}
%\bmsection{Funding}
\subsection*{Funding}
The work is supported by Russian Science Foundation grant \#24-19-00461.

%\bmsection{Acknowledgments}
%No.

%\bmsection{Disclosures}
%The authors declare no conflicts of interest.

%\bmsection{Data Availability Statement}
%Data underlying the results presented in this paper are not publicly available at this time but may be obtained from the authors upon reasonable request.

%\end{backmatter}

\bibliographystyle{unsrt}
\bibliography{refs}
%\bibliographyfullrefs{refs}

\end{document}